\documentclass[a4paper,12pt]{article}
\usepackage{epsfig}
\usepackage[dvips,usenames]{color}
\usepackage{axodraw}
\usepackage{graphicx}

\newlength{\dinwidth}
\newlength{\dinmargin}
\begin{document}
\title{$B\to \pi \pi, K \pi$ Decays in the QCD Improved Factorization Approach}
\author{Taizo Muta$^a$,   Akio Sugamoto$^b$,  Mao-Zhi Yang$^a$, Ya-Dong Yang$^b$\\
\small{$a$ Physics Department, Hiroshima University, Higashi-Hiroshima, 
Hiroshima 739-8526, Japan}\\
\small{$b$ Physics Department, Ochanomizu University, 2-1-1 Otsuka, Bunkyo-ku, 
Tokyo 112-8610, Japan}\\
}
\maketitle

\begin{picture}(0,0)
       \put(335,250){HUPD-0002}
       \put(335,270) {OCHA-PP-154}
       \put(335,290){\bf hep-ph/0006022}
\end{picture}

\begin{abstract}
Motivated by the recent measurements, we investigate $B\to \pi \pi, K \pi$ 
decay modes in the framework of QCD improved factorization, 
 which was recently proposed by Beneke $et\, al$. We find that all
the measured branching ratios are 
well accommodated in the reasonable parameter space except for $B\to K^0 \pi^0 $. 
We also discuss in detail the strong penguin contributions and the 
${\cal O}(\alpha_s )$ corrections to the chirally enhanced terms.
We find that the  weak phase $\gamma$ lies in the region 
$120^{\circ}<\gamma<240^{\circ}$, 
which is mainly constrained by $B\to \pi^- \pi^+ $.
\end{abstract}

\section{Introduction}
It is well known that the theoretical  description of nonleptonic  $B$ decays is an 
extremely outstanding challenge, due to the nonperturbative nature of both initial and 
final mesons. A good understanding of the B nonleptonic decays, or at least a reliable 
estimation, is the prerequisite for extracting meaningful implications from experimental 
data and for testing the SM. In past years, some achievements have been performed toward 
the goal, for example, in Ref. \cite{BSW, zep, gr}.

Recently, Beneke, Buchalla, Neubert and Sachrajda \cite{beneke} have presented a
promising factorization formula for the charmless nonleptonic B decays. The basic 
object in  the calculation of B charmless nonleptonic decays is the hadronic matrix
 element 
$\langle M_{1}(p_{1}) M_{2}(p_{2}){\mid}{\cal O}_i {\mid}B(p)\rangle$, where 
${\cal O}_{i}$ is the effective operator inducing the decay, $M_{1}$ is the final 
meson absorbing the light spectator quark from B meson, and $M_{2}$ is another light
meson flying fast from the b quark decay point as implied by ${\cal O}_i$.  The 
light spectator quark is translated softly to $M_{1}$ and this effect would be taken 
to the nonperturbative form factor $F_{1,2}^{B\to M_{1}}$ unless it undergoes a hard 
interaction. The quark pair, forming  $M_{2}$, ejected from b decay point carrying 
large energy of order of $m_b$ will involve hard interaction,  since soft gluon with 
momentum  of order $\Lambda_{QCD}$  will decouple from the quark pair at leading order
in $\Lambda_{QCD}/m_{b}$ in the heavy quark limit.  The essence of the argument 
of \cite{beneke} can be summarized by the improved factorization formula
\begin{eqnarray}
 \langle M_{1}(p_{1}) M_{2}(p_{2}){\mid}{\cal O}_i {\mid}B(p)\rangle &=& 
F^{B\to M_{1}}(M_{2}^{2}) \int_{0}^{1}dx T_{i}^{I}(x)\phi_{M_2}(x) \nonumber \\ 
&+& 
\int_{0}^{1}dxdydz T_{i}^{II}(x,y,z)\phi_{M_1}(x)\phi_{M_2}(y)\phi_{B}(z) ,  
\label{fac}
\end{eqnarray}
where $\phi_{P}(x)$ are the P meson's light-cone distribution amplitudes(DA). The  hard 
amplitudes $T_{i}^{I, II}$ can be perturbatively expanded in $\alpha_{s}(m_{b})$ and 
can be obtained from the calculations of the diagrams in Fig.1. It is interesting to
note that $T_{i}^{I}$ would be unity and $T_{i}^{II}$ would be absent at zeroth order
of $\alpha_{s}$ in the formula of Eq.(\ref{fac}), then the ${\it Naive}$  
${\it Factorization}$ would be reproduced.  Another consequence of Eq.(\ref{fac}) is 
that the final
 state interactions
may be computable and appear to be the imaginary part of the hard scattering amplitudes.
   
In this work,  we extend the formalism to $\bar{B}\to K \pi$ decays and recalculate 
$\bar{B} \to \pi \pi$ decays with electroweak penguin contributions. We also present 
detailed discussions about the strong penguin contributions and therefore we obtain the  
corrections to the chiral enhanced terms, which are found free of infrared divergence. 
We point out that there 
is large cancellation between the strong penguin hard scattering amplitudes and 
its contributions are small. Prospects of observing CP violation in those
decay modes are also discussed.

\section{Calculations} 

First we begin with the weak effective Hamiltonian $H_{eff}$
for the $\Delta B=1$ transitions as\cite{buras}
\begin{equation}
\label{heff}
{\cal H}_{eff}
=\frac{G_{F}}{\sqrt{2}} \left[ V_{ub} V_{uq}^*
\left(\sum_{i=1}^{2}
C_{i}O_{i}^{u}+  
\sum_{i=3}^{10}
C_{i} \, O_i + C_g O_g \right)+ 
V_{cb} V_{cq}^*
\left(\sum_{i=1}^{2}
C_{i}O_{i}^{c}+\sum_{i=3}^{10}
C_{i}O_{i}+C_{g}O_g \right)
 \right].
\end{equation}
For convenience, we list 
 below the operators in  ${\cal H}_{eff}$ for $b \to q$:
\begin{equation}\begin{array}{llllll}
O_1^{u} & = &\bar q_\alpha\gamma^\mu L u_\alpha\cdot \bar 
u_\beta\gamma_\mu L b_\beta\ ,
&O_2^{u} & = &  \bar q_\alpha\gamma^\mu L u_\beta\cdot \bar 
u_\beta\gamma_\mu L b_\alpha\ , \\
O_1^{c} & = &\bar q_\alpha\gamma^\mu L c_\alpha\cdot \bar 
c_\beta\gamma_\mu L b_\beta\ ,
&O_2^{c} & = &  \bar q_\alpha\gamma^\mu L c_\beta\cdot \bar 
c_\beta\gamma_\mu L b_\alpha\ , \\
O_3 & = & \bar q_\alpha\gamma^\mu L b_\alpha\cdot \sum_{q'}\bar
 q_\beta'\gamma_\mu L q_\beta'\ ,   &
O_4 & = & \bar q_\alpha\gamma^\mu L b_\beta\cdot \sum_{q'}\bar 
q_\beta'\gamma_\mu L q_\alpha'\ , \\
O_5 & = & \bar q_\alpha\gamma^\mu L b_\alpha\cdot \sum_{q'}\bar 
q_\beta'\gamma_\mu R q_\beta'\ ,   &
O_6 & = & \bar q_\alpha\gamma^\mu L b_\beta\cdot \sum_{q'}\bar 
q_\beta'\gamma_\mu R q_\alpha'\ , \\
O_7 & = & \frac{3}{2}\bar q_\alpha\gamma^\mu L b_\alpha\cdot 
\sum_{q'}e_{q'}\bar q_\beta'\gamma_\mu R q_\beta'\ ,   &
O_8 & = & \frac{3}{2}\bar q_\alpha\gamma^\mu L b_\beta\cdot 
\sum_{q'}e_{q'}\bar q_\beta'\gamma_\mu R q_\alpha'\ , \\
O_9 & = & \frac{3}{2}\bar q_\alpha\gamma^\mu L b_\alpha\cdot 
\sum_{q'}e_{q'}\bar q_\beta'\gamma_\mu L q_\beta'\ ,   &
O_{10} & = & \frac{3}{2}\bar q_\alpha\gamma^\mu L b_\beta\cdot 
\sum_{q'}e_{q'}\bar q_\beta'\gamma_\mu L q_\alpha'\ ,\\
O_g &=& (g_s/8\pi^{2}) \, m_b \, \bar{d}_{\alpha} \, \sigma^{\mu \nu}
      \, R  \, (\lambda^A_{\alpha \beta}/2) \,b_{\beta}
      \ G^A_{\mu \nu}~. 
\label{operators}
\end{array}
\end{equation}
Here $q=d,s$  and $(q'\epsilon\{u,d,s,c,b\})$. 
$\alpha$ and $\beta$ are the $SU(3)$ color indices and 
$\lambda^A_{\alpha \beta}$, $A=1,...,8$ are the Gell-Mann matrices;
$L$ and $R$ are the left- and right-handed projection operators with
$L=(1 - \gamma_5)$, $R= (1 + \gamma_5)$,
and $G^A_{\mu \nu}$ denotes the gluonic field strength tensor.
The Wilson coefficients evaluated at $\mu=m_b$ scale are\cite{buras}
\begin{equation}
\begin{array}{ll}
	C_1= 1.082, &
	C_2= -0.185,\\
	C_3=  0.014, &
	C_4= -0.035,\\
	C_5=  0.009, &
	C_6= -0.041,\\
	C_7= -0.002/137,&
	C_8=  0.054/137,\\
	C_9= -1.292/137,&
	C_{10}= -0.262/137,\\
        C_g=-0.143.&
\end{array}\label{ci}
\end{equation}

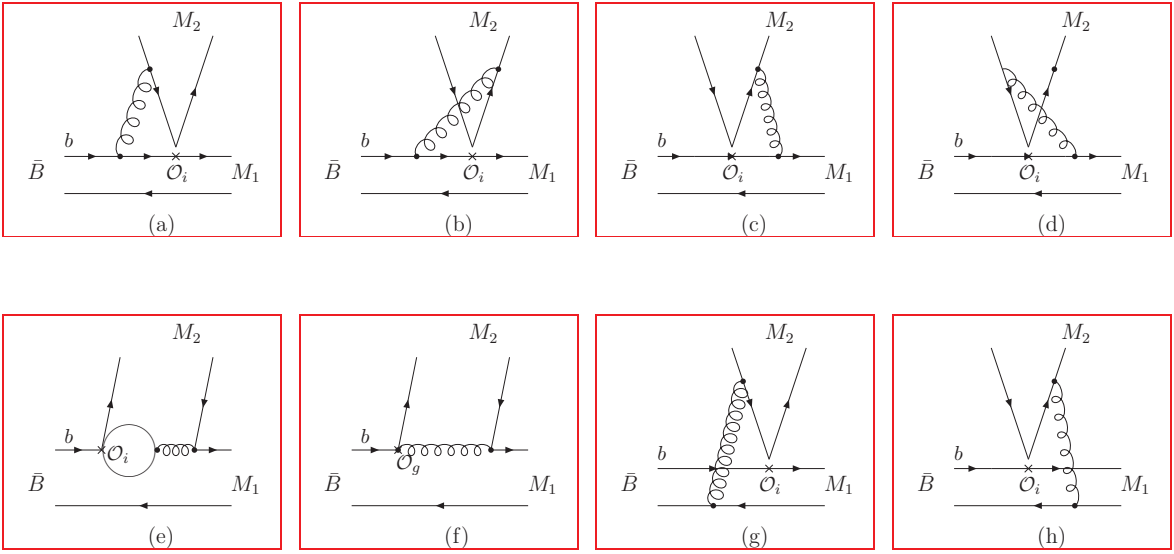
\begin{figure}[htbp]
 \scalebox{0.7}{
 {\color{Red}
 \fbox{\color{Black}
   \begin{picture}(140,120)(-30,0)
    \ArrowLine(0,40)(30,40)
    \ArrowLine(30,40)(60,40)
    \ArrowLine(60,40)(90,40)  
    \ArrowLine(90,20)(0,20)
    \Gluon(30,40)(46,87){4}{4} \Vertex(30,40){1.5} \Vertex(46,87){1.5} 
    \Line(58,42)(62,38)
    \Line(58,38)(62,42)
    \ArrowLine(40,105)(60,45)
    \ArrowLine(60,45)(80,105)
    \put(-20,28){$\bar{B}$}
    \put(90,28){$M_{1}$}
    \put(58,110){$M_{2}$}
    \put(55,28){\small{${\cal O}_i $}}
    \put(0,45){\small{$b$}}
    \put(45,0){(a)}
 \end{picture}
 }}}
 \scalebox{0.7}{
 {\color{Red}
 \fbox{\color{Black}
   \begin{picture}(140,120)(-30,0)
    \ArrowLine(0,40)(30,40)
    \ArrowLine(30,40)(60,40)
    \ArrowLine(60,40)(90,40)  
    \ArrowLine(90,20)(0,20)
    \Gluon(30,40)(74,87){4}{6} \Vertex(30,40){1.5} \Vertex(74,87){1.5} 
    \Line(58,42)(62,38)
    \Line(58,38)(62,42)
    \ArrowLine(40,105)(60,45)
    \ArrowLine(60,45)(80,105)
    \put(-20,28){$\bar{B}$}
    \put(90,28){$M_{1}$}
    \put(58,110){$M_{2}$}
    \put(55,28){\small{${\cal O}_i $}}
    \put(0,45){\small{$b$}}
    \put(45,0){(b)}
 \end{picture}
 }}}
 \scalebox{0.7}{
 {\color{Red}
 \fbox{\color{Black}
   \begin{picture}(140,120)(-30,0)
    \ArrowLine(0,40)(20,40)
    \ArrowLine(20,40)(60,40)
    \ArrowLine(60,40)(90,40)  
    \ArrowLine(90,20)(0,20)
    \Gluon(54,87)(65,40){3}{6} \Vertex(65,40){1.5} \Vertex(54,87){1.5} 
    \Line(38,42)(42,38)
    \Line(38,38)(42,42)
    \ArrowLine(20,105)(40,45)
    \ArrowLine(40,45)(60,105)
    \put(-20,28){$\bar{B}$}
    \put(90,28){$M_{1}$}
    \put(58,110){$M_{2}$}
    \put(35,28){\small{${\cal O}_i $}}
    \put(0,45){\small{$b$}}
    \put(45,0){(c)}
 \end{picture}
 }}}
 \scalebox{0.7}{
 {\color{Red}
 \fbox{\color{Black}
   \begin{picture}(140,120)(-30,0)
    \ArrowLine(0,40)(20,40)
    \ArrowLine(20,40)(60,40)
    \ArrowLine(60,40)(90,40)  
    \ArrowLine(90,20)(0,20)
    \Gluon(26,87)(65,40){3}{6} \Vertex(65,40){1.5} \Vertex(54,87){1.5} 
    \Line(38,42)(42,38)
    \Line(38,38)(42,42)
    \ArrowLine(20,105)(40,45)
    \ArrowLine(40,45)(60,105)
    \put(-20,28){$\bar{B}$}
    \put(90,28){$M_{1}$}
    \put(58,110){$M_{2}$}
    \put(35,28){\small{${\cal O}_i $}}
    \put(0,45){\small{$b$}}
    \put(45,0){(d)}
 \end{picture}
 }}}

\vspace{1cm} 
\scalebox{0.7}{
 {\color{Red}
 \fbox{\color{Black}
   \begin{picture}(140,120)(-30,0)
    \ArrowLine(90,20)(-5,20)
    \ArrowLine(-5,50)(20,50)
    \ArrowLine(20,50)(30,100)
    \ArrowLine(70,50)(90,50)  
    \ArrowLine(80,100)(70,50)
   \put(35,50){\circle{27}}
    \Gluon(50,50)(70,50){3}{3} \Vertex(50,50){1.5} \Vertex(70,50){1.5} 
    \Line(18,52)(22,48)
    \Line(18,48)(22,52)
    \put(-20,28){$\bar{B}$}
    \put(90,28){$M_{1}$}
    \put(58,110){$M_{2}$}
    \put(23, 45){\small{${\cal O}_i $}}
    \put(0,54){\small{$b$}}
    \put(45,0){(e)}
 \end{picture}
 }}}
\scalebox{0.7}{
 {\color{Red}
 \fbox{\color{Black}
   \begin{picture}(140,120)(-30,0)
    \ArrowLine(90,20)(-5,20)
    \ArrowLine(-5,50)(20,50)
    \ArrowLine(20,50)(30,100)
    \ArrowLine(70,50)(90,50)  
    \ArrowLine(80,100)(70,50)
    \Gluon(20,50)(70,50){3}{6} \Vertex(20,50){1.5} \Vertex(70,50){1.5} 
    \Line(18,52)(22,48)
    \Line(18,48)(22,52)
    \put(-20,28){$\bar{B}$}
    \put(90,28){$M_{1}$}
    \put(58,110){$M_{2}$}
    \put(19, 40){\small{${\cal O}_g $}}
    \put(0,54){\small{$b$}}
    \put(45,0){(f)}
 \end{picture}
 }}}
\scalebox{0.7}{
 {\color{Red}
 \fbox{\color{Black}
   \begin{picture}(140,120)(-30,0)
    \ArrowLine(0,40)(60,40)
    \ArrowLine(60,40)(90,40)  
    \ArrowLine(90,20)(0,20)
    \Gluon(30,20)(46,87){4}{10} \Vertex(30,20){1.5} \Vertex(46,87){1.5} 
    \Line(58,42)(62,38)
    \Line(58,38)(62,42)
    \ArrowLine(40,105)(60,45)
    \ArrowLine(60,45)(80,105)
    \put(-20,28){$\bar{B}$}
    \put(90,28){$M_{1}$}
    \put(58,110){$M_{2}$}
    \put(55,28){\small{${\cal O}_i $}}
    \put(0,45){\small{$b$}}
    \put(45,0){(g)}
 \end{picture}
 }}}
\scalebox{0.7}{
 {\color{Red}
 \fbox{\color{Black}
   \begin{picture}(140,120)(-30,0)
    \ArrowLine(0,40)(20,40)
    \ArrowLine(20,40)(90,40)  
    \ArrowLine(90,20)(0,20)
    \Gluon(54,87)(65,20){3}{6} \Vertex(65,20){1.5} \Vertex(54,87){1.5} 
    \Line(38,42)(42,38)
    \Line(38,38)(42,42)
    \ArrowLine(20,105)(40,45)
    \ArrowLine(40,45)(60,105)
    \put(-20,28){$\bar{B}$}
    \put(90,28){$M_{1}$}
    \put(58,110){$M_{2}$}
    \put(35,28){\small{${\cal O}_i $}}
    \put(0,45){\small{$b$}}
    \put(45,0){(h)}
 \end{picture}
 }}}
\caption{Order $\alpha_s$ corrections to the hard scattering kernels
 $T_{i}^I$ (a, b, c, d, e, f) and $T_{i}^{II}$ (g,h) }
\end{figure}

After direct calculations, we get the hard scattering for the decay modes 
listed as follows
\begin{equation}
   \begin{array}{ll}
   {\cal T}_{p}=&\frac{G_F}{\sqrt{2}}\sum_{p=u,c}V_{pq}^{*}V_{pb}\left[
           a_{1}^{p}(\bar{q}\gamma_{\mu}Lu)\otimes (\bar{u}\gamma^{\mu}Lb)
        +a_{2}^{p}(\bar{u}\gamma_{\mu}Lu)\otimes (\bar{q}\gamma^{\mu}Lb) \right.\\
     &+a_{3}^{p}(\bar{q'}\gamma_{\mu}Lq')\otimes (\bar{q}\gamma^{\mu}Lb)
     + a_{4}^{p}(\bar{q}\gamma_{\mu}Lq')\otimes (\bar{q'}\gamma^{\mu}Lb)
     + a_{5}^{p}(\bar{q'}\gamma_{\mu}Rq')\otimes (\bar{q}\gamma^{\mu}Lb) \\
    &+a_{6}^{p}(-2)(\bar{q}Rq')\otimes (\bar{q'}Lb)
   + a_{7}^{p} \frac{3}{2}e_{q'} (\bar{q'}\gamma_{\mu}Rq')\otimes 
             (\bar{q}\gamma^{\mu}Lb) 
   + (-2) (a_{8}^{p} \frac{3}{2}e_{q'} +a_{8a}) (\bar{q}Rq')\otimes 
             (\bar{q'}Lb) \\
 &+a_{9}^{p}\frac{3}{2}e_{q'}(\bar{q'}\gamma_{\mu}Lq')\otimes 
            (\bar{q}\gamma^{\mu}Lb)
  + (a_{10}^{p}\frac{3}{2}e_{q'}+a_{10a}^{p})
         (\bar{q}\gamma_{\mu}Lq')\otimes (\bar{q'}\gamma^{\mu}Lb)],\\
\end{array}
\label{tp}
\end{equation}
where the symbol $\otimes$ denotes 
$\langle M_1 M_2|j_2\otimes j_1|B \rangle \equiv \langle M_2|j_2|0\rangle 
  \langle M_1|j_1|B\rangle $. The 
 effective $a_{i}^p$'s which contain next-to-leadingorder(NLO)  coefficients and 
${\cal O}(\alpha_{s})$  hard scattering corrections are found to be
\begin{eqnarray}
 a_{1,2}^{c}&=&0, \quad a_{i}^{c}=a_{i}^{u}, i=3,5,7,8,9,10,8a,10a. \nonumber \\
 a_{1}^{u}&=&C_{1}+\frac{C_{2}}{N}+\frac{\alpha_{s}}{4\pi}\frac{C_F}{N}C_{2}F_{M_2},\nonumber \\
 a_{2}^{u}&=&C_{2}+\frac{C_{1}}{N}+\frac{\alpha_{s}}{4\pi}\frac{C_F}{N}C_{1}F_{M_{2}},\nonumber \\
 a_{3}^{u}&=&C_{3}+\frac{C_{4}}{N}+\frac{\alpha_{s}}{4\pi}\frac{C_F}{N}C_{4}F_{M_{2}}, \nonumber \\
 a_{4}^{p}&=&C_{4}+\frac{C_{3}}{N}+\frac{\alpha_{s}}{4\pi}\frac{C_F}{N}
\left[C_{3}(F_{M_{2}}+G_{M_{2}}(s_{q})+G_{M_{2}}(s_{b}))+C_{1}G_{M_{2}}(s_{p}) 
\right. \nonumber \\
& &\left. +(C_{4}+C_{6})\sum_{f=u}^{b}G_{M_{2}}(s_{f})+C_{g}G_{M_{2},g}\right],\nonumber\\
a_{5}^{u}&=&C_{5}+\frac{C_{6}}{N}+\frac{\alpha_{s}}{4\pi}\frac{C_F}{N}C_{6}(-F_{M_{2}}-12),
\nonumber \\
a_{6}^{p}&=&C_{6}+\frac{C_{5}}{N}
+\frac{\alpha_{s}}{4\pi}\frac{C_F}{N} 
\left[C_{1}G^{\prime}_{M_{2}}(s_{p})+C_{3}(G^{\prime}_{M_{2}}(s_{q})
+G^{\prime}_{M_{2}}(s_{b}))
+(C_{4}+C_{6})\sum_{f=u}^{b}G^{\prime}_{M_{2}}(s_{f})
+C_{g} G^{\prime}_{M_{2},g}\right], \nonumber \\
a_{7}^{u}&=&C_{7}+\frac{C_{8}}{N}+\frac{\alpha_{s}}{4\pi}\frac{C_F}{N}C_{8}(-F_{M_{2}}-12),
\nonumber \\
a_{8}^{p}&=&C_{8}+\frac{C_{7}}{N}, \nonumber \\
a_{8a}^{p}&=&\frac{\alpha_{s}}{4\pi}\frac{C_F}{N}
\left[(C_{8}+C_{10})\sum_{f=u}^{b}\frac{3}{2}e_{f}G^{\prime}_{M_{2}}(s_{f})
+C_{9}\frac{3}{2}(e_{q}G^{\prime}_{M_{2}}(s_{q})+e_{b}G^{\prime}_{M_{2}}(s_{b}))
\right], \nonumber \\
 a_{9}^{u}&=&C_{9}+\frac{C_{10}}{N}+\frac{\alpha_{s}}{4\pi}\frac{C_F}{N}C_{10}F_{M_{2}},
\nonumber \\
 a_{10}^{u}&=&C_{10}+\frac{C_{9}}{N}+\frac{\alpha_{s}}{4\pi}\frac{C_F}{N}C_{9}F_{M_{2}},
\nonumber \\
a_{10a}^{p}&=&\frac{\alpha_{s}}{4\pi}\frac{C_F}{N}
\left[(C_{8}+C_{10})\frac{3}{2}\sum_{f=u}^{b}e_{f}G_{M_{2}}(s_{f})
+C_{9}\frac{3}{2}(e_{q}G_{M_{2}}(s_{q})+e_{b}G_{M_{2}}(s_{b}))
\right], 
\label{aeff}
\end{eqnarray}
where $q=d, s.\quad$  $q'=u, d, s$ and $f=u, d, s, c, b$.  $C_{F}=(N^2 -1)/(2N)$ and $N=3$ 
is the  number of colors.  The internal quark mass in the penguin diagrams enters as 
$s_{f}=m_{f}^2/m_{b}^{2}$. $\bar{x}=1-x$ and $\bar{u}=1-u$.  
\begin{eqnarray}  
F_{M_{2}}&=&-12\ln\frac{\mu}{m_b } -18 + f_{M_{2}}^{I} + f_{M_{2}}^{II}, 
 \label{func0} \\
f_{M_{2}}^{I}&=&\int_{0}^{1}dx g(x)\phi_{M_{2}}(x), {\hskip 15mm}
g(x)= 3\frac{1-2x}{1-x}\ln x -3i\pi, \nonumber \\  
f_{M_{2}}^{II}&=&\frac{4\pi^2 }{N}\frac{f_{M_1}f_B }{ f_{+}^{B\to M_{1}}(0)
M_{B}^2 } \int_{0}^{1}dz\frac{\phi_{B}(z)}{z} 
\int_{0}^{1}dx\frac{\phi_{M_1}(x)}{x}
 \int_{0}^{1}dy\frac{\phi_{M_2}(y)}{y}, \\
G_{M_{2}, g} &=& -\int_{0}^{1}dx \frac{2}{\bar{x}}\phi_{M_2}(x), \\
G_{M_2}(s_{q}) &=&  \frac{2}{3} - \frac{4}{3}\ln\frac{\mu}{m_b} + 
4\int_{0}^{1}dx \phi_{M_2}(x) \int_{0}^{1}du\quad u\bar{u}\ln
\left[s_{q} -u\bar{u}\bar{x} -i\epsilon \right],\\
G^{\prime}_{M_{2}, g} &=& -\int_{0}^{1}dx \frac{3}{2}\phi^{0}_{M_2}(x)=-\frac{3}{2}, \\
G^{\prime}_{M_2}(s_{q}) &=&  \frac{1}{3} - \ln\frac{\mu}{m_b} + 
3\int_{0}^{1}dx \phi^{0}_{M_2}(x) \int_{0}^{1}du\quad u\bar{u}\ln
\left[s_{q} -u\bar{u}\bar{x} -i\epsilon \right],
\label{func}
\end{eqnarray}
where $\phi(x)$ and $\phi^{0}(x)$ are the meson's  leading-twist DA and 
twist-3 DA respectively. 
It should be noted that we have included   ${\cal O}(\alpha_s )$ corrections to $a_6$
in Eq.(\ref{aeff}). Although the $a_6$ term in Eq.(\ref{tp}) 
is formally $1/M_b$ suppressed,
it is chirally enhanced by $\mu_{P}=M^2_{P}/(m_q +m_{\bar{q}'})$  and known to 
be important to interpret the CELO\cite{cleo} measurement. As a result the 
 ${\cal O}(\alpha_s )$ correction to $a_6$ would be the most important one among 
the corrections to $a_i$. We see that there are logarithm terms $ln\mu/m_b$
appearing in Eqs.(\ref{func0}$\sim$\ref{func}), which is the result of one
loop integration. If the scale $\mu$ is chosen small, the logarithm would be
large and has to be resummed by using the renormalization group method. In this paper
we choose $\mu=m_b$, then the logarithm disappeared and the resummation is not
necessary. As a result, the effective coefficients $a_i^p$'s are obtained to the
order of $\alpha_s(m_b)$ corrections ( see also in Ref.\cite{flei}).  
  
We realize that the contribution of  the strong penguins depicted in fig.1.(e) and (f) 
 to $a_6$ could be reliably estimated  without IR divergence.
As an example, we show the contribution of
Fig.1(f) in the following. With the assignment of the  vertex 
$\delta_{\alpha\beta}if_{M_2}\mu_{M_2}\gamma_5 \phi^{0}(x)/4N_c $ to 
$M_2 $ and its constituents, we  can get the hard amplitudes of Fig.1.(f) as
\begin{eqnarray}
H_{f} &\sim&  if_{M_{2}}\mu_{M_2}\frac{\alpha_{s}}{4\pi}\frac{C_F}{N}
   \int_{0}^{1}dx \phi^{0}(x) \frac{3(1-x)m_{b}^2 }{k^2 }
\bar{q}_i \gamma_{\mu}(1-\gamma_{5}) b_i  \nonumber \\
&\sim& \bar{q}_i \gamma_{\mu}(1-\gamma_{5}) b_i  \int_{0}^{1}dx \phi^{0}(x).    
\label{maf}
\end{eqnarray}
We can see that the end point IR divergence in $1/k^2 (k^2 =(1-x)m_{b}^2 )$ is canceled
by the term $(1-x)$ in the numerator and the amplitude is finite. For the amplitude of 
Fig.1.e, it is easy to note that the denominator $k^2$ of the gluon propagator is canceled
by the quark loop and the integration of $\int_{0}^{1} dx G(s_f)$ is also finite itself.
However, if all the external quarks are treated as free qurks at first, IR divergence 
will appear. In the case of free quarks, one can get the hard amplitudes of Fig.1.(f) as
\begin{eqnarray}
H_{f} &\sim& \frac{m_{b}^2}{k^2} 
\bar{d_i} \gamma_{\mu}(1-\gamma_{5}) b_j 
 \bar{q_j}\gamma^{\mu}q_i    \nonumber \\
&\sim&\frac{m_b^2}{k^2} 
\left[\bar{d_i}\gamma_{\mu}(1-\gamma_{5})b_j 
\bar{q_j}\gamma^{\mu}(1-\gamma_{5})q_i +
  \bar{d_i} \gamma_{\mu}(1-\gamma_{5}) b_j 
 \bar{q_j}\gamma^{\mu}(1+\gamma_{5}) q_i \right].    
\label{mef}
\end{eqnarray}
At this stage the quark pair $\bar{q}d$  is in  color-singlet configuration. 
After $Fierz$ rearrangement, one 
gets
 \begin{equation}
H_{f} \sim  \frac{m_{b}^2}{k^2}
\left[ \bar{d_i} \gamma_{\mu} (1-\gamma_{5}) q_i 
\otimes \bar{q_j}\gamma^{\mu}(1-\gamma_{5})b_j 
-2 \bar{d_i} (1+\gamma_{5}) q_i 
\otimes \bar{q_j}(1-\gamma_{5})b_j .  
\right]
\end{equation}
From the above equation we can see that Fig.1(f) contributes
to $a_{4}$ and  $a_{6}$ equally  and its contribution is IR divergent when 
$k^2 \to 0$ in free quark approach. Phenomenologically, one may have to treat 
$k^2$ as a parameter. In the framework employed here, the virtuality of the 
gluon is convoluted with the meson's DA.  Furthermore, The NLO strong penguin 
contributions to $a_4$ and $a_6$ terms are different.   
  
Finally, the chirally enhanced contribution of Fig.1(g) and (h) to $a_6$
is cancelled between them. One can easily see this cancellation by putting
both the leading-twist DA and twist-3 DA $\phi(x)$ and $\phi^0(x)$ to 
Fig.1(g) and (h) and calculate these two diagrams. Because $\phi^0(x)$ gives
the chirally enhanced contributions, one can easily see that these contributions
are cancelled. 

With Eqs.(\ref{tp})and (\ref{aeff}), we can write down the amplitudes of
$B\to \pi \pi$ and $K\pi$ decays
 \begin{eqnarray}
{\cal M}(\bar{B_{d}^0} \to\pi^{+}\pi^{-})&=&
\frac{G_{F}}{\sqrt{2}}if_{\pi}(M_{B}^{2}-M_{\pi}^{2})F^{B\to\pi}(0){\mid}\lambda 
V_{cb}{\mid}
\biggl\{ R_{b}e^{-i\gamma}\left[a_{1}^{u}+a_{4}^{u}
+a_{10}^{u}+a_{10a}^{u} \right. \biggr. \nonumber \\
&&\biggl. \left. +R_{\pi^-}(a_{6}^{u}+a_{8}^{u}+a_{8a})\right] 
 -\left[ a_{4}^{c}+a_{10}^{c}+a_{10a}^{c}+R_{\pi}(a_{6}^{c}+a_{8}^{c}
+a_{8a})\right]
\biggr\} , \\
{\cal M}(\bar{B_{d}^0} \to\pi^{0}\pi^{0})&=&
\frac{G_{F}}{\sqrt{2}}if_{\pi}(M_{B}^{2}-M_{\pi}^{2})F^{B\to\pi}(0){\mid}\lambda 
V_{cb}{\mid}\times   \nonumber \\ 
&&\biggl\{ R_{b}e^{-i\gamma}\left[-a_{2}^{u}+a_{4}^{u}
+\frac{3}{2}a_{7}^{u}-\frac{3}{2}a_{9}^{u}-\frac{1}{2}a_{10}^{u}
+a_{10a}^{u}       
 +R_{\pi^0}(a_{6}^{u}-\frac{1}{2}a_{8}^{u}+a_{8a})\right] \biggr. \nonumber \\
&&\biggl. -\left[a_{4}^{c}+\frac{3}{2}a_{7}^{c}
-\frac{3}{2}a_{9}^{c}-\frac{1}{2}a_{10}^{c}
+a_{10a}^{c}+R_{\pi^{0}}(a_{6}^{c}-\frac{1}{2}a_{8}^{c} +a_{8a})\right]
\biggr\}, \\
{\cal M}(\bar{B_{u}^{-}}\to\pi^{0}\pi^{-})&=&
\frac{G_{F}}{2}if_{\pi}(M_{B}^{2}-M_{\pi}^{2})F^{B\to\pi}(0){\mid}\lambda 
V_{cb}{\mid}
\biggl\{ R_{b}e^{-i\gamma}\left[ a_{1}^{u}+a_{2}^{u} \right. \biggr. \nonumber \\
&& \biggl. \left. +\frac{3}{2}(-a_{7}^{u}+R_{\pi}a_{8}^{u}+a_{9}^{u}+a_{10}^{2})\right] 
-\frac{3}{2}\left[ -a_{7}^{c}+R_{\pi^0}a_{8}^{c}+a_{9}^{c}+a_{10}^{c}) \right]
\biggr\},\\
{\cal M}(\bar{B_{d}^0} \to\bar{K^0}\pi^{0})&=&
\frac{G_{F}}{2}if_{\pi}(M_{B}^{2}-M_{K}^{2})F^{B\to K}(0)(1-\lambda^2 )
{\mid} V_{cb}{\mid} \times \nonumber \\
& &\biggl\{
  R_{b}^{\prime}e^{-i\gamma}\left[a_{2}^{u}
-\frac{3}{2}\left( a_{7}^{u} - a_{9}^{u}\right)\right] 
-\frac{3}{2}\left( a_{7}^{u} - a_{9}^{u}\right)
\biggr\}    \nonumber \\
&&-\frac{G_{F}}{2}if_{K}(M_{B}^{2}-M_{\pi}^{2})F^{B\to \pi}(0)(1-\lambda^2 )
{\mid} V_{cb}{\mid} \times \nonumber \\
& &\left\{ 
  R_{b}^{\prime}e^{-i\gamma}\left[ -a_{4}^{u}
-R_{K}\left( a_{6}^{u}-\frac{1}{2}a_{8}^{u} +a_{8a}\right)   
+\frac{1}{2}a_{10}^{u} -a_{10a}^{u}
\right]  \right. \nonumber \\
&&\left.  +\left[ -a_{4}^{c}
-R_{K}\left( a_{6}^{c}-\frac{1}{2}a_{8}^{c}+a_{8a}\right)   
+\frac{1}{2}a_{10}^{c} -a_{10a}^{c}
\right] 
\right\}. \label{kopo}\\
{\cal M}(\bar{B_{d}^0} \to K^{-}\pi^{+})&=&
\frac{G_{F}}{\sqrt{2}}if_{\pi}(M_{B}^{2}-M_{K}^{2})F^{B\to K}(0)(1-\lambda^2 )
{\mid} V_{cb}{\mid} \times \nonumber \\
& &\biggl\{
  R_{b}^{\prime}e^{-i\gamma}\left[a_{1}^{u} + a_{4}^{u} 
+R_{K}\left( a_{6}^{u}+a_{8}^{u}+a_{8a}\right) + a_{10}^{u} + a_{10a}^{u} \right] 
\biggr. \nonumber \\
& &\biggl. \left[ a_{4}^{c} 
+R_{K}\left( a_{6}^{c}+a_{8}^{c}\right) + a_{10}^{c} + a_{10a}^{c} \right]
\biggr\}. \\
{\cal M}(\bar{B_{u}^{-}} \to K^{-}\pi^{0})&=&
\frac{G_{F}}{2}if_{K}(M_{B}^{2}-M_{\pi}^{2})F^{B\to \pi}(0)(1-\lambda^2 )
{\mid} V_{cb}{\mid} \times \nonumber \\
& &\biggl\{
  R_{b}^{\prime}e^{-i\gamma}\left[a_{1}^{u} + a_{4}^{u} 
+R_{K}\left( a_{6}^{u}+a_{8}^{u}+a_{8a}\right) + a_{10}^{u} + a_{10a}^{u} \right] 
\biggr. \nonumber \\
& &\biggl. \left[ a_{4}^{c} 
+R_{K}\left( a_{6}^{c}+a_{8}^{c}+a_{8a}\right) + a_{10}^{c} + a_{10a}^{c} \right]
\biggr\} \nonumber \\
&+&\frac{G_{F}}{2}if_{\pi}(M_{B}^{2}-M_{K}^{2})F^{B\to K}(0)(1-\frac{\lambda^{2}}{2} )
{\mid} V_{cb}{\mid} \times \nonumber \\
& &\biggl\{
  R_{b}^{\prime}e^{-i\gamma}\left[a_{2}^{u} +
  \frac{3}{2}\left( a_{9}^{u} - a_{7}^{u} \right)\right]
   + \frac{3}{2}\left( a_{9}^{c} - a_{7}^{c} \right)
\biggr\}.  \\
 {\cal M} (\bar{B_{u}^{-}} \to \bar{K^{0}}\pi^{-})&=&
\frac{G_{F}}{\sqrt{2}}if_{K}(M_{B}^{2}-M_{\pi}^{2})
F^{B\to \pi}(0)(1-\frac{\lambda^{2}}{2} )
{\mid} V_{cb}{\mid} \times \nonumber \\
& &\biggl\{
  R_{b}^{\prime}e^{-i\gamma}\left[a_{4}^{u}
+R_{K}\left( a_{6}^{u}-\frac{1}{2}a_{8}^{u}+a_{8a}\right)
-\frac{1}{2}a_{10}^{u} + a_{10a}^{u} \right]
\biggr. \nonumber \\
&&\biggl. +\left[a_{4}^{c} + 
+R_{K}\left( a_{6}^{c}-\frac{1}{2}a_{8}^{c}+a_{8a}\right)
-\frac{1}{2}a_{10}^{c} + a_{10a}^{c} \right]
\biggr\}
\end{eqnarray}
Where $R_b =\frac{1-\lambda^{2}/2}{\lambda}{\mid} \frac{V_{ub}}{V_{cb}}{\mid}$ and
 $R_{b}^{\prime} =\frac{\lambda}{1-\lambda^{2}/2} {\mid}\frac{V_{ub}}{V_{cb}}{\mid}$.  
$V_{cb}, V_{ud}$ and $V_{us}$ are chosen to be real and $\gamma$ is the phase of 
$V^{*}_{ub}$. $\lambda=|V_{us}|=0.2196$. $R_{P}=2\mu_{P}$.
  
\section{Numerical calculations and discussions of results}

In the numerical calculations we use \cite{pdg}
$$f_\pi = 0.133{ GeV},~~f_K =0.158GeV, ~~f_B = 0.180GeV,$$
$$\tau(B^+)=1.65\times 10^{-12}s, ~~~\tau(B^0)=1.56\times 10^{-12}s,$$
$$ M_B = 5.2792 { GeV},~~~~ M_b =4.8GeV, ~~~~M_c =1.4 GeV,$$
$$m_u=4.0 MeV, ~~M_d=9.0MeV, ~~M_s =80MeV.$$ 
For the leading-twist DA $\phi(x)$ and the twist-3 DA $\phi^{0}(x)$ of K and $\pi$,  
we use the well known asymptotic form of these DA\cite{chz,braun} 
\begin{equation}
\phi_{\pi,K}(x) =  6 x (1-x), \quad \phi^{0}_{\pi,K}(x) =1.
\label{phipp}
\end{equation}
For $B$ meson, the wave function is chosen as that used in \cite{kls,luy}
\begin{equation}
\phi_B(x) = N_{B} x^2(1-x)^2 \mathrm{exp} \left
 [ -\frac{M_B^2\ x^2}{2 \omega_{B}^2} \right],
\label{bwav}
\end{equation}
with  $\omega_{B}=0.4$ GeV, and $N_{B}$ is the 
normalization constant to make $\int_{0}^{1} dx \phi_{B}(x) =1$. Here the decay
constant in the wave function has been factored out. So the wave function can be
normalized to $1$. It is also necessary to 
note that $\phi_{B}(x)$  is strongly peaked around $x=0.1$. This character is 
consistent with the observation of Heavy Quark Effective Theory that the 
wave function should be peaked around $\Lambda_{QCD}/M_{B}$. 
With such choice, 
we find
\begin{equation}
\int_{0}^{1} dx \frac{\phi_{B}(x)}{x}=11.15,  
\end{equation}
which is near to the argument\cite{beneke} in which
$\int_{0}^{1} dx \phi_{B}(x)/x =M_{B}/{\lambda_{B}}=17.56$ with $\lambda_{B}=0.3GeV$.
 We have used the unitarity of the CKM matrix 
$V_{uq}^* V_{ub} + V_{cq}^* V_{cb} + V_{tq}^* V_{tb}=0$  to decompose the amplitudes 
into terms containing $V_{uq}^* V_{ub}$ and  $V_{cq}^* V_{cb}$, and 

\begin{equation}\begin{array}{ll}
|V_{ud}|=1-\lambda^{2}/2,& |V_{ub}/V_{cb}|=0.085\pm 0.02 \\
|V_{cb}|=0.0395\pm 0.0017& |V_{us}|=\lambda=0.2196.
\end{array}
\end{equation}
We leave the CKM angle $\gamma$ as a free parameter. For the form factors,
we use $F^{B\to \pi}(0)=0.3$  and  $F^{B\to K}(0)=1.13F^{B\to \pi}(0)$.

Numerical values for $a_{i}^{p}(\pi\pi)$ and $a_{i}^{p}(\pi K)$  are 
presented in Table 1. It should be noted that $a_{i}(K\pi)$ are generally
different to $a_{i}(\pi\pi)$ and also change from case to case due to 
$f_{M_2}^{II}$ in the formulas of $a_{i}$, where $M_{2}$ could be $K$
or $\pi$. However, with our choice of parameters 
\begin{equation}  
\frac{f_{\pi}}{F^{B\to \pi}(0)}\simeq \frac{f_{K}}{F^{B\to K}(0)}, \nonumber
\end{equation}
and the same DAs $\phi_{K,\pi}(x)$, the 
$a_{i}(K\pi)\simeq a_{i}(\pi\pi)$.
From Table.1, we can find that all $a_{i}^{p}$ 
develop strong phases due to hard strong scattering. 
Our $a_{2}$ is very different from that of \cite{ali2, cheng2} in both real and
imaginary part because of the contribution of Fig.1.(g), and (h). So, theoretical 
predictions for the decays dominated by $a_2$ may be very different between 
$Naive$ $Factorization$ approach and QCD improved factorization approach. 
Numerically,  we find that the ${\cal O}(\alpha_s )$ strong penguin 
contributions which collected  in $a_{4}$ and $a_{6}$ are small
because of the large cancellation between Fig.1.e and Fig.1.f. 
In detail, the strong penguin contributions to $a_{4}$ and $a_6$ are 
\begin{eqnarray}
a_{4pen}^{p}&=&\frac{\alpha_{s}}{4\pi}\frac{C_F}{N} 
\left[C_{1}G_{M_{2}}(s_{p})+C_{3}(G_{M_{2}}(s_{q})+G_{M_{2}}(s_{b}))
+(C_{4}+C_{6})\sum_{f=u}^{b}G_{M_{2}}(s_{f})+C_gG_{M_{2},g}\right]  \nonumber \\
 &=& \frac{\alpha_{s}}{4\pi}\frac{C_F}{N} \times \left\{ 
\begin{array}{ll}
(-0.780-1.744i)+(0.858), &\quad p=u, \\
(-1.473-0.529i)+(0.858),  &\quad p=c. 
\end{array} 
\right.
\end{eqnarray}

\begin{eqnarray}
a_{6pen}^{p}&=&\frac{\alpha_{s}}{4\pi}\frac{C_F}{N} 
\left[C_{1}G^{\prime}_{M_{2}}(s_{p})+C_{3}(G^{\prime}{M_{2}}(s_{q})
+G^{\prime}_{M_{2}}(s_{b}))
+(C_{4}+C_{6})\sum_{f=u}^{b}G^{\prime}_{M_{2}}(s_{f})+C_gG^{\prime}_{M_{2},g}\right]  \nonumber \\
 &=& \frac{\alpha_{s}}{4\pi}\frac{C_F}{N} \times \left\{ 
\begin{array}{ll}
(-0.780-1.299i)+(0.2145), &\quad p=u, \\
(-1.095-0.510i)+(0.2145),  &\quad p=c. 
\end{array} 
\right.
\end{eqnarray}
where the numbers in the brakets are the contibutions of Fig.1.e and Fig.1.f respectively.
The cancellation in $a_6$ is weaker than that in $a_4$, since the contribution of Fig.1.f 
to $a_6$ is small. 
The other diagrams will dominate the ${\cal O}(\alpha_s)$ hard scattering amplitudes.

\begin{table}[htbp]
\caption{The QCD coefficients $a_{i}^{p}$ at NLO for renormalization scale
$\mu=m_b$ ( in units of $10^{-4}$ for $a_{3,...,} a_{10}$). 
Results from different references are shown for comparison. }
\begin{center}
	\begin{tabular}{ccccc}
		\hline  
            \hline
         & ours  & \cite{beneke}& \cite{ali2}&\cite{cheng2} \\
$a_{1}$  & 1.042+0.014i  & 1.038+0.018i & 1.05    & 1.46     \\
$a_{2}$  & 0.046-0.082i  & 0.082-0.080i   & 0.053 & 0.24     \\
$a_{3}$ & 65.2+26.8i  & 40+20i   & 48      & 72     \\
\hline
$a_{4}^{u}$ &-314-152i   & -290-150i   &-439-77i& -383-121i     \\
$a_{4}^{c}$ &-370-54i   & -340-80i   &     &      \\
\hline
$a_{5} $ &-55.7-31.4i   &-50-20i    &-45     & -27     \\
\hline  
$a_{6}^u$ &-380+(-46-106i)   & -380   &-575-77i  & -435-121i     \\
$a_{6}^c$ &-380+(-71-41i)   & -380  &     &      \\
\hline
$a_{7}$  &1.25+0.3i   &    & 0.5-1.3i    & -0.89-2.73i    \\
\hline
$a_{8}$  &3.8+(-0.1-0.5i) &    & 4.6-0.4i  & 3.3-0.91i     \\
\hline
$a_{9} $ &-98.4+1.47I &    & -94-1.3i    &-93.9-2.7i      \\
$a_{10}$  &-39.3+7.23i  &    & -14-0.4i    &0.32-0.90i  \\    		
\hline
\hline
	\end{tabular}
\end{center}
\label{tab}
\end{table}

Now it is time to discuss branching ratios and CP asymmetries of $B\to K\pi$
and $B\to \pi\pi$ in the QCD improved factorization approach. The branching
ratio is given by
\begin{equation}
Br(B\to K\pi,\pi\pi)=\tau_B/(16\pi m_B)|{\cal M(B\to K\pi,\pi\pi)}|^2 s,
\end{equation}
where $s=1/2$ for $B\to \pi^0\pi^0$ mode, and $s=1$ for the other decay modes.
For the charged $B$ meson decays, the direct CP asymmetry parameter is defined as
\begin{equation}
A^{dir}_{CP}=\frac{|{\cal M}(B^+\to f)|^2-|{\cal M}(B^-\to \bar{f})|^2}
                      {|{\cal M}(B^+\to f)|^2+|{\cal M}(B^-\to \bar{f})|^2}.
\label{dircp}
\end{equation}
For the neutral B decaying into CP eigenstate $f$, i.e., $f=\bar{f}$,
the effects of $B^0-\bar{B^0}$ mixing should be taken
into account in studying CP asymmetry. Thus the CP asymmetry is time dependent,
which is given by\cite{Grcp}
\begin{equation}
A_{CP}(t)=A^{dir}_{CP}cos(\Delta mt)-
             \frac{2Im(\lambda_{CP})}{1+|\lambda_{CP}|^2} sin(\Delta mt),
\label{mixcp}
\end{equation}
where $\Delta m$ is the mass difference of the two mass eigenstates of neutral B
mesons, and $A^{dir}_{CP}$ is the direct CP asymmetry defined in 
eq.(\ref{dircp}) with replacement of $B^+\to B^0$ and $B^-\to \bar{B^0}$, respectively. 
The parameter $\lambda_{CP}$ is given by
\begin{equation}
\lambda_{CP}=\frac{V_{tb}^*V_{td}\langle f|H_{eff}|\bar{B^0}\rangle}
                  {V_{tb}V_{td}^*\langle f|H_{eff}|B^0\rangle}.
\end{equation}

With the above parameters and formulae, we get the branching ratios
\begin{eqnarray}
{\cal BR}(\bar{B_{d}^0} \to\pi^{+}\pi^{-})&=& 
7.55\times10^{-6} |e^{-i\gamma} + 0.18  e^{i8.0^\circ }|^2 , \nonumber \\ 
{\cal BR}(\bar{B_{d}^0} \to\pi^{0}\pi^{0})&=&4.3\times10^{-8}|  e^{-i\gamma} 
+ 1.19 e^{-i132^\circ }|^2 , \nonumber \\  
{\cal BR}(B_{u}^{-}\to\pi^{0}\pi^{-})&=&4.73\times10^{-6}|  e^{-i\gamma} 
 + 0.05  e^{-i0.1^\circ }|^2 ,\nonumber \\
{\cal BR}(\bar{B_{d}^0} \to\bar{K^0}\pi^{0})&=&4.06\times10^{-9}|  e^{-i\gamma} 
 + 31.9   e^{i34^\circ } |^2 ,\label{phas} \\  
{\cal BR}(\bar{B_{d}^0} \to K^{-}\pi^{+})&=&5.12\times10^{-7}|  e^{-i\gamma} 
 + 5.23e^{-i172^\circ } |^2 , \nonumber \\ 
{\cal BR}(B_{u}^{-} \to K^{-}\pi^{0})&=&2.91\times10^{-7}|  e^{-i\gamma} 
 + 5.78  e^{-i168^\circ } |^2 ,\nonumber \\
 {\cal BR} (B_{u}^{-} \to \bar{K^{0}}\pi^{-})&=&4.08\times10^{-9}|  e^{-i\gamma}
 + 55.1  e^{-i11.3^\circ } |^2 . \nonumber 
\end{eqnarray}

If we generally express eq.(\ref{phas}) as ${\cal BR}=A(e^{-i\gamma}+ae^{-i\delta})$,
then the direct CP asymmetry in eq.(\ref{dircp}) can be relevantly expressed as
\begin{equation}
  A^{dir}_{CP}=\frac{2a sin\gamma}{1+a^2+2a cos\delta cos\gamma}.
\end{equation}
Using the above equation, the numerical results for the direct CP asymmetry are
obtained
\begin{eqnarray}
A^{dir}_{CP}(B\to\pi^+\pi^-)&=&\frac{5.0\%}{1.03+0.36cos\gamma }sin\gamma,\nonumber \\
A^{dir}_{CP}(B\to\pi^0\pi^0)&=&-\frac{1.77}{2.42-1.59cos\gamma }sin\gamma,\nonumber \\
A^{dir}_{CP}(B\to\pi^0\pi^\mp)&=&-1.7\times 10^{-4}sin\gamma,\nonumber \\
A^{dir}_{CP}(B\to K^0\pi^0)&=&3.5\%sin\gamma,\label{cpnum} \\
A^{dir}_{CP}(B\to K^\mp\pi^\pm)&=&-\frac{1.46}{28.4-10.4cos\gamma }sin\gamma,\nonumber \\
A^{dir}_{CP}(B\to K^\mp\pi^0)&=&-\frac{2.40}{34.4-11.3cos\gamma }sin\gamma,\nonumber \\
A^{dir}_{CP}(B\to K^0\pi^\mp)&=&-0.7\%sin\gamma.\nonumber 
\end{eqnarray}
 
As is shown in Eq.(\ref{phas}), the strong phases are different by decay channels. 
We can also see from eq.(\ref{cpnum})
that the direct CP violation in $B\to \pi^{0}\pi^{\mp}$   
is neglectably small. The direct CP violation in 
$B \to\pi^{+}\pi^{-}$, $\pi^{0}K^{\mp}$, $K^{0}\pi^{0}$, $K^{\mp}\pi^{0}$
 and $K^{\mp}\pi^{\pm}$ are only at a few percentage
level. The large CP violation effect may be expected in $B \to \pi^{0}\pi^{0}$ decays.
However, it would be remained undetectable before the running of  
the next generation B factories, 
for example, LHCB, due to its very small branching ratios$(\sim 10^{-7})$ and its 
two neutral final states.  

Recently, the CLEO collaboration has made first observation of the decay modes
$B\to \pi^{+}\pi^{-}$, $B\to K^{0}\pi^{0}$ and $B\to K^{\pm}\pi^{0}$ and also 
updated the decay modes $B\to K^{\pm}\pi^{\mp}$ and $B\to K^{0}\pi^{\pm}$ as 
follows\cite{cleo}
\begin{eqnarray} 
{\cal BR}(B_{d} \to\pi^{+}\pi^{-})&=& 
(4.3^{+1.6}_{-1.4}\pm 0.5)\times10^{-6}, \nonumber \\ 
{\cal BR}(B_{u}\to\pi^{0}\pi^{\pm})&<& 12.7 \times10^{-6}, \nonumber \\ 
{\cal BR}(B_{d} \to K^0 \pi^{0})&=&
(14.6^{+5.9+2.4}_{-5.1-3.3})\times10^{-6}, \nonumber \\ 
{\cal BR}(B_{d} \to K^{\pm}\pi^{\mp})&=&
(17.2^{+2.5}_{-2.4}\pm 1.2)\times10^{-6}, \\ 
{\cal BR}(B_{u} \to K^{\pm}\pi^{0})&=&
(11.6^{+3.0+1.4}_{-2.7-1.3})\times10^{-6}, \nonumber \\ 
 {\cal BR} (B_{u} \to K^{0}\pi^{\pm})&=&
(18.2^{+4.6}_{-4.0}\pm 1.6)\times10^{-6}, \nonumber  
\end{eqnarray}

\begin{figure}
\begin{tabular}{cc}
\scalebox{0.7}{\epsfig{file=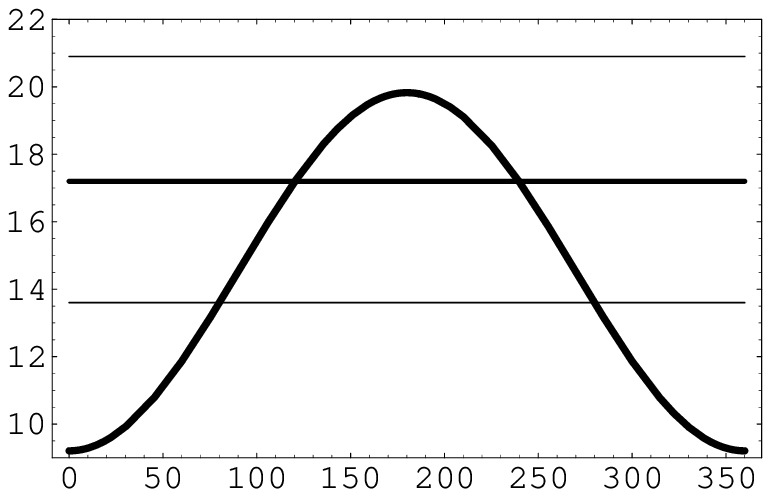}} & \scalebox{0.7}{\epsfig{file=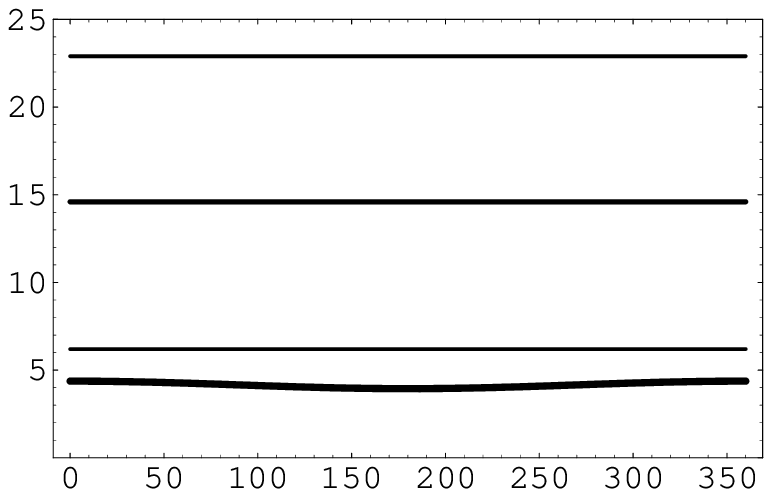}}  \\ 
\small{Fig.2.1, $Br(B \to K^{\mp} \pi^{\pm})$ vs $\gamma$} &
\small{Fig.2.2, $Br(B \to K^{0} \pi^{0})$ vs $\gamma$}     \\[4mm]
\scalebox{0.7}{\epsfig{file=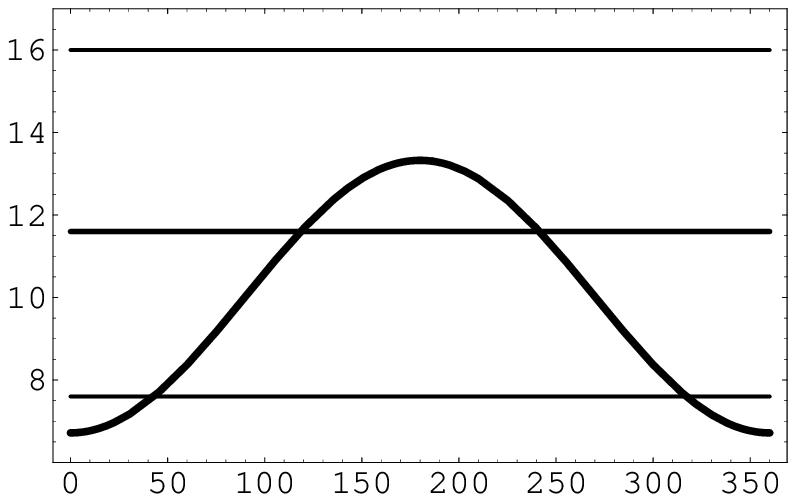}} & \scalebox{0.7}{\epsfig{file=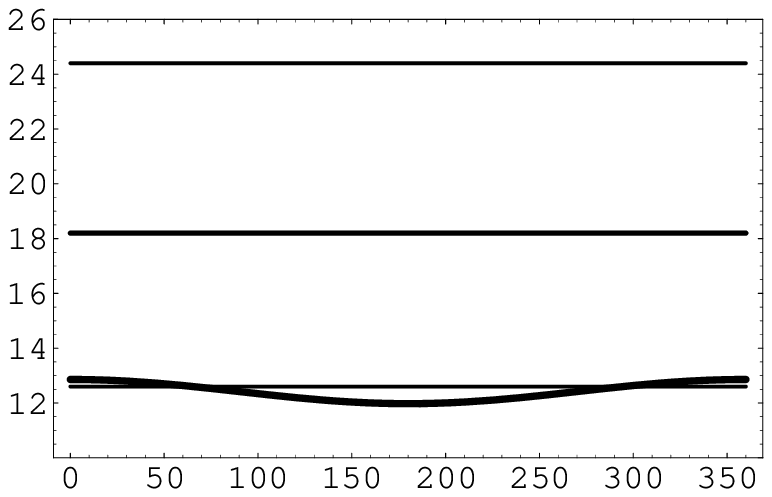}}  \\ 
\small{Fig.2.3, $Br(B \to K^{\mp} \pi^{0})$ vs $\gamma$} &
\small{Fig.2.4, $Br(B \to K^{0} \pi^{\mp})$ vs $\gamma$} \\[4mm]
\scalebox{0.7}{\epsfig{file=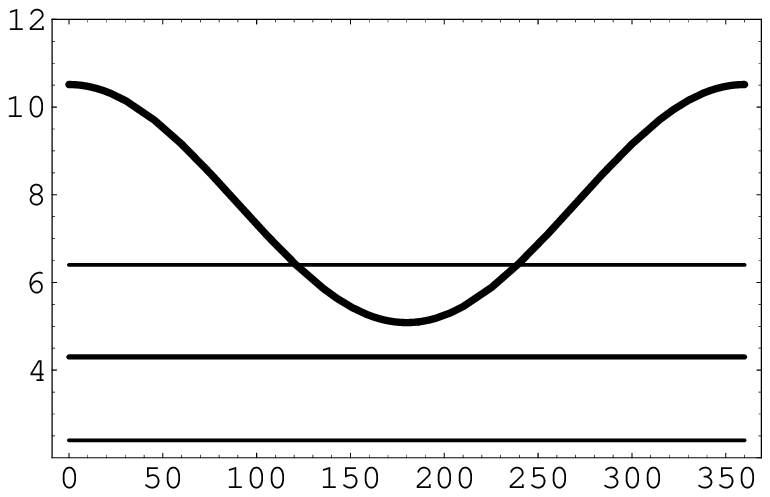}} & \scalebox{0.7}{\epsfig{file=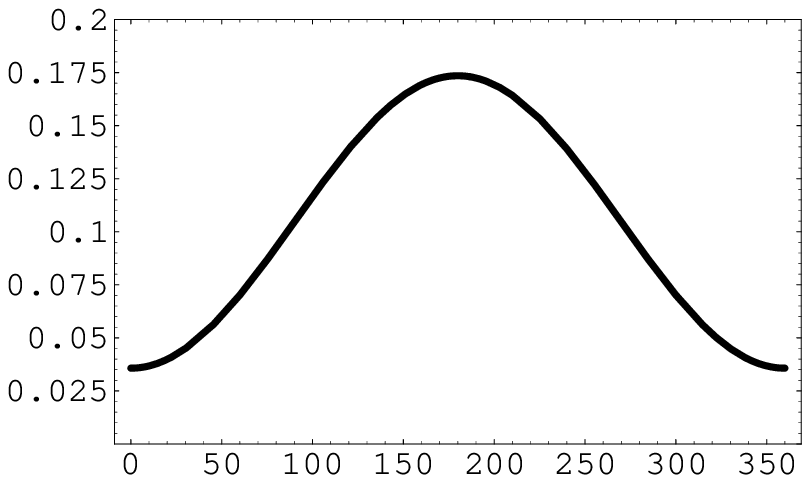}}  \\ 
\small{Fig.2.5, $Br(B \to \pi^{\mp} \pi^{\pm})$ vs $\gamma$} &
\small{Fig.2.6, $Br(B \to \pi^{0} \pi^{0})$ vs $\gamma$} \\[4mm]
\scalebox{0.7}{\epsfig{file=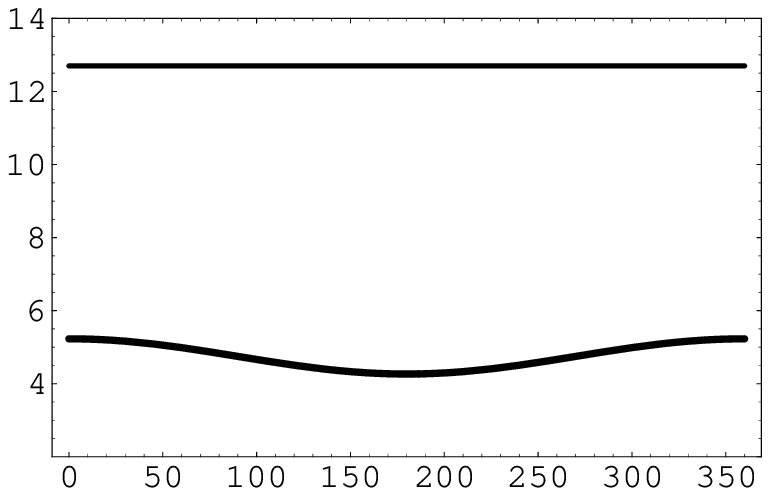}} & \\ 
\small{Fig.2.7, $Br(B \to \pi^{\mp} \pi^{0})$ vs $\gamma$} &
\end{tabular}
\caption{\small{CP-averaged $BR( B\to \pi\pi, K\pi)$ as a function of $\gamma$ are shown as 
 curves for $F^{B\to\pi}=0.3$ and $|V_{ub}/V_{cb}|=0.08$ (in units of $10^{-6}$). 
The BR measured by CLEO 
Collaboration are shown by horizontal solid lines. The thicker solid lines are 
its center values, thin lines are its error bars or the upper limit.}}    

\end{figure}

To compare with the data, we plot the CP averaged branching ratios for those 
modes as a function of $\gamma$ in Fig.2. Our results are plotted as curves
and the CELO data are displayed as horizontal lines ( thicker lines for center
value, thin lines represent error bars at 2$\sigma$ level). The horizontal line in 
Fig.2.7 is the upper limit of the decay mode.  
 
We find that the observed branching ratios of those decay modes can be well accommodated 
within the QCD improved factorization approach of Ref\cite{beneke} except the decay 
mode $B\to K^0 \pi^0 $. As shown in Eq.(\ref{kopo}), the first term with $F^{B\to K}$
and the second term with $F^{B\to \pi}$
are $disconstructive$  which  reduces the amplitude of 
$M(B\to K^0 \pi^0 )$ much smaller than that of other  $B\to \pi K$ decays.     
As it is argued in Ref\cite{beneke, beneke1}, in the present theoretical framework, 
the final state interactions are computable
and identical to the imaginary part of the amplitude which is generated by the 
 hard scattering  amplitudes. In this paper, we find the 
strong phase appears not large enough to change the two sub-amplitudes of $M(B\to K^0 \pi^0)$ 
to be $constructive$. Our results agree with that in 
Ref.\cite{ali2, cheng1, hou, cheng3} 
where the decay rate of  $B\to K^0 \pi^0 $ is also estimated to 
be small.

The CLEO observations have motivated many theoretical studies of those decay 
modes using  different approaches\cite{kls, luy, cheng1, hou, gronau}. 
In Refs.\cite{hou, desh, he}, it is suggested that $\gamma>90^{\circ}$ 
is required to 
interpret the CLEO data. However, the global CKM fit has given the constraint
$\gamma< 90^{\circ}$  at $99.6\%$ C.L.\cite{cara}. The comparison between our 
results and CLEO data\cite{cleo} implies  
$120^{\circ}<\gamma <240^{\circ}$ which arises from the constraint by 
$Br(B\to \pi^- \pi^+ )$. The observed $Br(B\to \pi^- \pi^+ )$ is smaller than 
many theoretical expectations. Negative $cos\gamma$ is needed to suppress the
theoretical estimations as it is suggested in Ref.\cite{hou}.  The decay
rate of $B\to \pi^- \pi^+ $  can be also suppressed by using smaller 
form factor $F^{B\to \pi}(0) $ and/or smaller $|V_{ub}/V_{cb}|$. However, it would
be very hard to account for the large decay rates of $B\to K\pi$ modes in this case.
For those reasons, it might be difficult to solve the controversy between the 
global CKM fit and the model-dependent constraints from the charmless decays
$B\to K\pi, \pi\pi$ within the QCD improved factorization approach.

\section{Summary}

We have studied 
$B \to K^{\mp} \pi^{\pm}$ 
$B \to K^{0} \pi^{0}$ 
$B \to K^{\mp} \pi^{0}$ 
$B \to K^{0} \pi^{\mp}$ 
$B \to \pi^{\mp} \pi^{\pm}$ 
$B \to \pi^{0} \pi^{0}$ and
$B \to \pi^{\mp} \pi^{0}$ 
decays, in QCD improved factorization approach. 

The strong penguin contributions (Fig.1.e,f) are discussed in detail and found
to be small because of the cancellations between them. The most important power
corrections to these chiral enhanced terms(i.e., $a_6$) are identified and found
to be free of infrared divergence. With the choice of twist-3  DA $\phi^{0}_p(x)=1$,
the $a_6$ gets a large imaginary part and its real part is enhanced by $10\sim20{\%}$.
The other NLO coefficients $a_i$ also acquire complex  phases from the hard scattering
as depicted by $Fig1.(a)\sim(e)$ which are shown by the function 
$g(x)$ and $G(s,x)$ in
Eq.(\ref{func}). We can see that $g(x)$ is a new source of strong phase 
besides $G(s,x)$ of the well known BSS mechanism\cite{bss}. Compared to the $Naive$ 
$factorization$, 
the strong phases 
are estimated reliably without the arbitrariness of gluon virtuality $k^2$  within
 the QCD improved 
factorization  formalism\cite{beneke}.      
The strong phase due to the hard scattering  in the decay modes are found to vary 
from $0^{\circ}$ to $172^{\circ}$, depending on the decay mode. In the decays 
$B\to \pi^0 \pi^0$, $K^{\pm}\pi^{\mp}$ and  $K^{\pm}\pi^{0}$, the strong phase are 
found to be as large as $100^{\circ}< \delta <180^{\circ}$. In other decay modes,
the strong phases are rather small.
 
 The predicted branching ratios of $B\to \pi K$  
and $B \to \pi^{\mp} \pi^{\pm}$ decay modes are in good agreement with
 the experimental measurement by the CLEO Collaboration 
except the decay $B\to K^0 \pi^0$.   The most serious constraint on 
the weak angle $\gamma$ comes from the small experimental value of  
$Br(B\to \pi^- \pi^+ )$ which implies $120^{\circ}< \gamma <240^{\circ}$.
We found that it is hard to  solve the controversy between the constraints on 
$\gamma$ from the global CKM fit and the estimations of the charmless 
decays $B\to K\pi, \pi\pi$.
The CP violation effects in  $B\to \pi^{0}\pi^{\mp}$   
is neglectably small. The direct CP violation effects in 
$B \to\pi^{+}\pi^{-}$, $\pi^{0}K^{\mp}$, $K^{0}\pi^{0}$, $K^{\mp}\pi^{0}$
 and $K^{\mp}\pi^{\pm}$ are only at a few percentage
level. The large CP violation effect may be expected in $B \to \pi^{0}\pi^{0}$ decays.

$\it Note \, added$: After finishing this work, we find Ref.\cite{du} also 
discussed $B\to K\pi$ and $\pi\pi$ decays with a similar method, and 
Ref.\cite{kl} compared different approaches. \par

${\bf Addendum}$ After the paper was sent  for publishing, the BARBAR Collaboration 
report their measurment of branching ratios for charmless B decays to charged pions
and kions\cite{barbar}: 
$B(B^0 \to \pi^{\pm} \pi^{\mp})=(9.3^{+2.6+1.2}_{-2.3-1.4})\times 10^{-6}$ 
and $B(B^0 \to K^{\pm}\pi^{\mp} )=(12.5^{+3.0+1.3}_{-2.6-1.7})\times 10^{-6}$. 
Our predictions agree with the BARBAR data very well. 
We note that $positive$ $\cos\gamma$ is favored if the BARBAR data is taken as guide.

\section*{Acknowledgments}

We acknowledge the Grant-in-Aid for Scientific Research on Priority Areas
(Physics of CP violation with contract number 09246105 and 1014028). 
Y.D.Yang and M.Z. Yang thank JSPS for support.

\begin{newpage}

\end{newpage}


\begin{thebibliography}{99}
\bibitem{BSW} M.Bauer, B.Stech, and M. Wirbel, Z. Phys. {\bf C29}, 637(1985),
Z. Phys. {\bf C34}, 103(1987).

\bibitem{zep}D.Zeppenfield, Z. Phys. {\bf C8}, 77(1981);
L.L.Chau,   Phys. Rev. {\bf D43}, 2176(1991).

\bibitem{gr}M.Gronau, J.Rosner and D. London, Phys. Rev. Lett. {\bf 73}, 21(1994) 

\bibitem{beneke}M. Beneke, G. Buchalla, M. Neubert, C.T. Sachrajda, 
Phys. Rev. Lett. 83, 1914 (1999). 

\bibitem{buras} For a review, see G. Buchalla, A.J. Buras,
M.E. Lautenbacher, Rev. Mod. Phys. 68, 1125 (1996).

\bibitem{cleo} D. Cronin-Hennessy, et al., CLEO Collaboration,
hep-ex/0001010.

\bibitem{flei} R. Fleischer, Z. Phys. {\bf C58}, 483 (1993).

\bibitem{pdg} Particle Data Group, Eur. Phys. J. C3, 1 (1998).

\bibitem{chz} V.L. Chernyak and A.R. Zhitinissky, Phys. Rep.{\bf 112},173(1983).

\bibitem{braun}V.M.braun and I.E.Filyanov, Z. Phys.{\bf C48}, 239(1990).

\bibitem{kls}Y.Y. Keum, H.-n. Li, A.I. Sanda, preprint KEK-TH-642, 
        NCKU-HEP-00-01,hep-ph/0004004;preprint NCKU-HEP-00-02,
        DPNU-00-14, hep-ph/0004173.
\bibitem{luy}C.D. L\"{u}, K. Ukai, M. Z. Yang, preprint HUPD-9924,
        DPNU-00-15, hep-ph/004213. 

\bibitem{ali2} A. Ali, G. Kramer and C.D. L\"u, Phys. Rev. D58, 094009
(1998); 

\bibitem{cheng2}Y.-H. Chen, H.-Y. Cheng, B. Tseng, K.-C. Yang, 
 Phys. Rev. {\bf D60}, 094014 (1999). 

\bibitem{Grcp}M. Gronau, Phys. Rev. Lett 63, 1451, (1989).

\bibitem{beneke1}M. Beneke, CERN-TH/99-319, hep-ph/9910505.

\bibitem{cheng1} H.Y.Cheng, Phys. Lett. {\bf B335}(1994)428; 
Phys. Lett. {\bf B395}(1997)345; H.Y.Cheng and B.Tseng, 
Phys. Rev. {\bf D58}(1998)094005; 
Y.H.Chen,  H.Y.Cheng and B.Tseng,   Phys. Rev. {\bf D59}(1999)074003. 

\bibitem{hou}W.S.Hou, and K.C.Yang, hep-ph/9908202, hep-ph/9911528;
W.S.Hou J.G.Smith and F.W\"urthwein, hep-ex/9910014.   

\bibitem{cheng3}H.-Y.Cheng nad K.-C. Yang, hep-ph/9910291. 

\bibitem{gronau} M.Gronau and J.L.Rosner, TECHNION-PH-99-33, EFI 99-40,
hep-ph/9909478,  Phys. Rev. {\bf D59}, 113002(1999).  

\bibitem{desh}N.G.Deshpande $et$ $al$., Phys. Rev. Lett. {\bf 82}, 2240(1999).   

\bibitem{he}X.G.He, W.S.Hou and K.C.Yang, Phys. Rev. Lett. {\bf 83}, 1100(1999).     

\bibitem{cara}F.Caravaglios, F.Parodi, P.Roudeau and A.Stocchi,
LAL 00-04, hep-ph/0002171. 

\bibitem{bss}M. Bander, D. Silverman, and A. Soni, Phys. Rev. Lett.
    43, 242(1979)  

\bibitem{du} D. Du, D. Yang and G. Zhu, hep-ph/0005006.

\bibitem{kl} Y.Y. Keum, and H.N. Li, hep-ph/0006001.
\bibitem{barbar} B. Aubert, et. al., the BARBAR Collaboration, BARBAR-CONF-00/14,
SLAC-PUB-8536. 
\end{thebibliography}
\end{document}